\global\def\draftcontrol{0}

   \def\versionno{ Calabi-Yau Heat Bound }
\catcode`\@=11

\expandafter\ifx\csname draftcontrol\endcsname\relax\global\def\draftcontrol{0} 
\fi 

{\count255=\time\divide\count255 by 60 
\xdef\hourmin{\number\count255} 
\multiply\count255 by-60\advance\count255 by\time 
\xdef\hourmin{\hourmin:\ifnum\count255<10 0\fi\the\count255}} 
\def\draftdate{\number\month/\number\day/\number\year\ \ \ \hourmin } 

\newcommand\makepapertitle{\par

  \begingroup 
    \renewcommand\thefootnote{\@fnsymbol\c@footnote}% 
    \def\@makefnmark{\rlap{\@textsuperscript{\normalfont\@thefnmark}}}% 
    \long\def\@makefntext##1{\parindent 1em\noindent 
            \hb@xt@1.8em{% 
                \hss\@textsuperscript{\normalfont\@thefnmark}}##1}% 
     \newpage 
     \global\@topnum\z@   % Prevents figures from going at top of page. 
     \@makepapertitle 
     \thispagestyle{empty}\@thanks 
  \endgroup 
  \setcounter{footnote}{0}% 
  \global\let\thanks\relax 
  \global\let\makepapertitle\relax 
  \global\let\@makepapertitle\relax 
  \global\let\@thanks\@empty 
  \global\let\@author\@empty 
  \global\let\@date\@empty 
  \global\let\@title\@empty 
  \global\let\title\relax 
  \global\let\author\relax 
  \global\let\date\relax 
  \global\let\and\relax 
  \def\version{\let\version\@version\@gobble} 
} 
\def\@makepapertitle{% 
  \newpage 
   \ifnum\draftcontrol=1 {} 
   \version\versionno 
   \vskip 5.5em% 
   \else 
   \hfill\hbox to 4.0cm {\parbox{5.5cm}{\@pubnum}\hss}% 
   \vskip 5.5em% 
   \fi 
   \begin{center}% 
   \let \footnote \thanks 
      {\hskip -0\textwidth \hbox to 1\textwidth% 
        {\centerline{\Large\bf{\noindent\@title}}}}% 
     \vskip 1.7em% 
     {\normalsize%\large 
       \lineskip .5em% 
       \begin{tabular}[t]{c}% 
         \@author 
       \end{tabular}\par}% 
     \vskip 1.2em% 
     {\@bstract}% 
     \end{center}% 
     \vfill
     \@date%
     \vfill%
     \noindent
     \rule{12em}{.02em}\par\noindent
     \@email%
   \par 
} 

\gdef\@pubnum{} 
\def\pubnum#1{% 
  \gdef\@pubnum{#1}} 

\gdef\@bstract{} 
\def\Abstract#1{% 
  \gdef\@bstract{% 
   \parbox{\textwidth-0pc}{% 
   \centerline{\bf Abstract}\penalty1000 
   \noindent%\abstractfont \baselineskip=12pt 
   \renewcommand\baselinestretch{1.0} 
   {#1}}} 
} 

\gdef\@email{}
\def\email#1{%
   \gdef\@email{%
   Email: {\tt #1}}
}

\def\ps@paper{\let\@mkboth\@gobbletwo% 
     \ifnum\draftcontrol=1 
        \def\@oddfoot{\hbox to \textwidth{\tiny \versionno \hfil\tiny\draftdate}% 
        \hskip -\textwidth \hbox to \textwidth{\hfil\rm\thepage\hfil}}% 
     \else\def\@oddfoot{\hbox to \textwidth{\hfil\rm\thepage\hfil}} 
     \fi 
     \let\@evenfoot\@oddfoot 
} 
\newenvironment{acknowledgments}{% 
\vskip 3.25ex 
\addcontentsline{toc}{section}{Acknowledgments}
\noindent {\bf Acknowledgments} 
} 

\def\@version#1{\ifnum\draftcontrol=1 
\typeout{}\typeout{#1}\typeout{} 
\vskip3mm\centerline{\hbox{\fbox{\normalsize{\tt DRAFT -- #1 -- } 
                   {\draftdate}}}}\vskip3mm 
\fi} 
\let\version\@version 
\long\def\eqlabel#1{\ifnum\draftcontrol=1 
                    \tag@false  % there are some problems with multline without this 
                    \tag*{(\theequation) \hbox to -0.2cm{\hspace{0cm}\small{#1}\hss}} 
                    \edef\@currentlabel{\theequation} 
                    \refstepcounter{equation}     
                    \ltx@label{#1}          % use old LaTeX \label instead of new definition 
                    \else 
                    \label{#1} 
                    \fi 
                    } 
\let\st@bibitem\@bibitem 
\let\st@lbibitem\@lbibitem 
\ifnum\draftcontrol=1 
  \def\@bibitem#1{% 
    \st@bibitem{#1}\a@@label{#1}\ignorespaces} 
  \def\@lbibitem[#1]#2{% 
    \st@lbibitem[#1]{#2}\a@@label{#2}\ignorespaces} 
  \def\a@@label#1{% 
    \gdef\a@lab{\smash{\normalfont\small#1}} 
    \ifvmode 
      \if@inlabel 
        \global\setbox\@labels\hbox{% 
          \llap{\a@lab\let\a@lab\relax 
                \kern\@totalleftmargin\kern\marginparsep}% 
          \box\@labels}% 
      \fi 
    \fi} 
\fi 
\documentclass[12pt,letterpaper]{article} 
\usepackage{amsmath,bm,amsfonts,amssymb,array,calc,amsthm,rotating}
\usepackage{epsfig,psfrag} 
\usepackage{graphicx}
\usepackage{color}
\usepackage[colorlinks=false]{hyperref}
\renewcommand\baselinestretch{1.25} 
\renewcommand\section{\@startsection {section}{1}{\z@}% 
                                   {-3.5ex \@plus -1ex \@minus -.2ex}% 
                                   {2.3ex \@plus.2ex}% 
                                   {\normalfont\large\bfseries}} 
\renewcommand\subsection{\@startsection{subsection}{2}{\z@}% 
                                   {-3.25ex\@plus -1ex \@minus -.2ex}% 
                                   {1.5ex \@plus .2ex}% 
                                   {\normalfont\normalsize\bfseries}} 
\renewcommand\subsubsection{\@startsection{subsubsection}{3}{\z@}% 
                                   {-3.25ex\@plus -1ex \@minus -.2ex}% 
                                   {1.5ex \@plus .2ex}% 
                                   {\normalfont\normalsize\it}} 
\renewcommand\paragraph{\@startsection{paragraph}{4}{\z@}% 
                                   {-3.25ex\@plus -1ex \@minus -.2ex}% 
                                   {1.5ex \@plus .2ex}% 
                                   {\normalfont\normalsize\bf}} 
\renewcommand\subparagraph{\@startsection{subparagraph}{5}{\z@}% 
                                   {-1.25ex\@plus -1ex \@minus -.2ex}% 
                                   {0ex \@plus .2ex}% 
                                   {\normalfont\normalsize\it}}

\numberwithin{equation}{section}

\long\def\@makecaption#1#2{%
  \vskip\abovecaptionskip
  \sbox\@tempboxa{{\bf #1:} #2}%
  \ifdim \wd\@tempboxa >\hsize
    {\small\bf #1:} {\small #2}\par
  \else
    \global \@minipagefalse
    \hb@xt@\hsize{\hfil\box\@tempboxa\hfil}%
  \fi
  \vskip\belowcaptionskip}
\renewcommand*\l@section[2]{%
  \ifnum \c@tocdepth >\z@
    \addpenalty\@secpenalty
    \addvspace{.5em \@plus\p@}%
    \setlength\@tempdima{1.5em}%
    \begingroup
      \parindent \z@ \rightskip \@pnumwidth
      \parfillskip -\@pnumwidth
      \leavevmode \bfseries
      \advance\leftskip\@tempdima
      \hskip -\leftskip
      #1\nobreak\hfil \nobreak\hb@xt@\@pnumwidth{\hss #2}\par
    \endgroup
  \fi}
\renewcommand*\l@subsection{\addvspace{.0em \@plus\p@}\@dottedtocline{2}{1.5em}{2.3em}}
\renewcommand*\l@subsubsection{\addvspace{-.2em \@plus\p@}\@dottedtocline{3}{3.8em}{3.2em}}

\def\hepth#1{\href{http://xxx.arxiv.org/abs/hep-th/#1}{{arXiv:hep-th/#1}}}

\def\mathdg#1{\href{http://xxx.arxiv.org/abs/math.DG/#1}{{arXiv:math.dg/#1}}}

\def\arxiv#1#2{\href{http://xxx.arxiv.org/abs/#1}{{arXiv:#1 [#2]}}}

\definecolor{refcol}{rgb}{0.2,0.2,0.8}
\definecolor{eqcol}{rgb}{.6,0,0}
\definecolor{purple}{cmyk}{0,1,0,0}

\gdef\@citecolor{refcol}
\gdef\@linkcolor{eqcol}
\def\colorlinkspurple{\gdef\@urlcolor{purple}}
\def\colorlinksblue{\gdef\@urlcolor{blue}}
\def\colorlinksred{\gdef\@urlcolor{red}}

\def\ie{{\it i.e.}}

\def\cf{{\it cf.}}

\def\revise#1       {\raisebox{-0em}{\rule{3pt}{1em}}% 
                     \marginpar{\raisebox{.5em}{\vrule width3pt\ 
                     \vrule width0pt height 0pt depth0.5em 
                     \hbox to 0cm{\hspace{0cm}{% 
                     \parbox[t]{4em}{\raggedright\footnotesize{#1}}}\hss}}}}

\def\calm         {{\cal M}} 
\def\caln         {{\cal N}}

\def\reals        {{\mathbb R}}

\def\ee           {{\it e}} 
\def\ii           {{\it i}} 
 
\def\tr           {{\rm Tr}}

\def\sqr#1#2{{\vcenter{\vbox{\hrule height.#2pt   
 \hbox{\vrule width.#2pt height#1pt \kern#1pt 
 \vrule width.#2pt}\hrule height.#2pt}}}}

\def\vol{\mathop{\rm vol}}

\catcode`\@=12

\usepackage{soul}
\usepackage[usenames,dvipsnames,svgnames,table,x11names]{xcolor}

\begin{document} 

%%% 
%%%%%% text starts here 
%%%%%%%%% 

\title{Bounding the Heat Trace of a Calabi-Yau Manifold}

\pubnum{arXiv:yymm.nnnn}
\date{June 2015}

\pubnum{
\arxiv{yymm.nnnnn}{hep-th}}
\date{June 2015}

\author{
Marc-Antoine Fiset$^{\dag}$, Johannes Walcher$^{\dag\ddag}$ \\[0.2cm]
\it $^{\dag}$ Department of Physics \\
\it McGill University, \it Montr\'eal, Qu\'ebec, Canada \\[.1cm]
\it $^{\ddag}$ Department of Physics, and Department of Mathematics and Statistics \\ 
\it McGill University,
\it Montr\'eal, Qu\'ebec, Canada}

\email{marc-antoine.fiset@mail.mcgill.ca, johannes.walcher@mcgill.ca}

\Abstract{
The SCHOK bound states that the number of marginal deformations of certain
two-dimensional conformal field theories is bounded linearly from above by the 
number of relevant operators. In conformal field theories defined via 
sigma models into Calabi-Yau manifolds, relevant operators 
%(``counter-factual Kaluza-Klein tachyons'') 
can be estimated, in the point-particle approximation, 
by the low-lying spectrum of the scalar Laplacian on the manifold. In the strict 
large volume limit, the standard asymptotic expansion of Weyl and 
Minakshisundaram-Pleijel diverges with the higher-order curvature invariants.
We propose that it would be sufficient to find an a priori uniform bound on the trace 
of the heat kernel for large but finite volume. As a first step in this direction, 
we then study the heat trace asymptotics, as well as the actual spectrum of the 
scalar Laplacian, in the vicinity of a conifold singularity. The eigenfunctions 
can be written in terms of confluent Heun functions, the analysis of
which gives evidence that regions of large curvature will not prevent the
existence of a bound of this type. This is also in line with general mathematical
expectations about spectral continuity for manifolds with conical singularities.
A sharper version of our results could, in combination 
with the SCHOK bound, provide a basis for a global restriction on the dimension 
of the moduli space of Calabi-Yau manifolds.
 }

\makepapertitle

%\body

\version\versionno

\vskip 1em

%\tableofcontents
\newpage

\tableofcontents

\section{Introduction}

Perturbation theory approximates the space of solutions of string theory by
the space of two-dimensional conformal field theories satisfying 
certain conditions on the chiral algebra, such as an appropriate central charge or 
extended supersymmetry, that allow a consistent coupling to two-dimensional 
gravity, and guarantee perturbative consistency and finiteness of the space-time 
theory \cite{chsw,fms}.

Non-perturbative quantum corrections \cite{dise} and dualities \cite{witten} change 
both the local and the global details of this picture of the space of string 
vacua, and will eventually stabilize even a non-supersymmetric vacuum 
\cite{kklt}. However, these modifications do not address the central issue 
of {\it finiteness} of the String Landscape \cite{susskind}, which, in the 
absence of other principles, would provide a viable approach to string 
phenomenology \cite{dode}.

In recent years, a revival of the conformal bootstrap program has led to
remarkable progress on a priori constraints on the {\it operator content} of certain 
types of conformal field theories. Of particular interest for the space of string 
vacua are the results of Hellerman and Schmidt-Colinet \cite{schh}, and Keller and 
Ooguri \cite{ook}, SCHOK: Exploiting only the constraint of modular invariance 
of the torus partition function (and, in the supersymmetric case, of the elliptic 
genus), these works have shown that there exists an intimate relationship between 
the number of marginal and relevant operators in two-dimensional (super-)conformal 
field theories. Besides its phenomenological relevance as a starting point for
phenomenological finiteness in string theory, 
a strict upper bound on the number of marginal operators 
of a SCFT of central charge $\hat c=3$ would imply the existence of an upper bound 
on the dimension of cohomology groups of Calabi-Yau threefolds (and hence their Euler 
number), which is a hopeful, but largely open, mathematical conjecture.

In the present note, we explore in some detail the geometric ramifications of the
SCHOK bound, as it applies to conformal field theories defined as the infrared
fixed point of a Calabi-Yau sigma model. On the one hand, marginal deformations 
of such a CFT are in one-to-one correspondence with certain elements of the 
cohomology groups of the underlying manifold \cite{lvw}. The associated harmonic
forms can be identified with the supersymmetric ground states in the Hilbert
space, which can be reached by spectral flow from the chiral ring spanned by the
marginal operators \cite{lvw}. This data also controls the massless spectrum of 
the space-time theory that results upon compactification of the ten-dimensional 
super-string on the Calabi-Yau.

On the other hand, relevant operators of the CFT correspond to states with negative
worldsheet energy, not far above the tachyonic ground state. In a supersymmetric 
string compactification, such states are eliminated by the GSO projection, and 
do not actually enter the space-time theory at all. Nevertheless,
they are present in the conformal field theory {\it before} GSO projection, and,
by the SCHOK argument, allow some control over the space of marginal operators
that do survive the GSO projection, and hence, the massless spectrum.

An important feature of this situation is that while the number of marginal operators
is of cohomological nature (``BPS'') and hence does not vary over the smooth part
of the moduli space (corresponding to space-time theories without extra, non-perturbative, 
massless states), the number of relevant operators, and their
conformal dimensions, are moduli-dependent quantities. It is natural to ask
whether the freedom that results from this distinction can be exploited 
to yield additional information on the spectrum of such conformal field theories.

The main idea of the present paper goes back to a question that arose in \cite{pc}: Are
there any interesting constraints on the number of relevant operators that
depend on a geometric origin of the conformal field theory, but that are 
independent of other, topological data such as the number of marginal deformations?
What sorts of constraints on the massless spectrum can be derived from these results?
 
In the perturbative $\alpha'$- (large-volume/small-curvature) expansion of the sigma 
model, the lightest string states are those without any
oscillators excited, \ie, they involve only the variables describing the motion
of the center of mass of the string on the manifold. Therefore, the questions
about the number of relevant operators of the conformal field theory become
questions of classical spectral geometry. We claim two main results.

First, we will argue, following \cite{pc}, that an upper bound on the trace of 
the heat kernel of the Calabi-Yau manifold at large temperature (or small time), 
possibly together with bounds on qualitatively 
similar quantities, would be a sufficient constraint on the light spectrum 
of the resulting CFT to complement the SCHOK bound. Importantly, this bound
need {\it not} hold everywhere in the moduli space of Ricci-flat metrics 
(and we do not expect that it does), but
it should be uniform in the topology of the manifold, in a sense that we will
explain.

As far as we know, no bound of this nature is presently available in the spectral 
geometry literature. There exists, of course, a standard large temperature expansion 
of the trace of the heat kernel going back to Weyl. However, the asymptotic nature 
of this expansion makes it insufficient for bounding purposes: The expansion 
coefficients are integrals of local curvature invariants of higher and higher 
order, so that estimating the remainder requires rescaling the metric to smooth out
regions of large curvature. This, however, requires that the volume be large, and
implies that the leading Weyl term cannot be controlled in a uniform fashion.

To investigate this issue, we study the spectrum of the scalar Laplacian and the behaviour
of the heat trace in the regime in which the manifold develops a curvature singularity,
but without relying on the asymptotic expansion previously mentioned. For concreteness, 
we focus on the approach to the conifold singularity of a Calabi-Yau
threefold \cite{conifold} ``from the resolved side'', but the structure of the 
argument will make it clear that more general singularities should not be very 
different. We formulate the problem in terms of spectral continuity under 
the confluence of Heun eigenfunctions. The high order WKB analysis of these solutions 
suggests an explanation of the smooth behaviour observed in the high curvature limit.

Our second result is then the conclusion that not only do regions of large curvature not 
prevent the existence of a uniform bound, but that in fact degenerations 
of the manifold could be a useful starting point to obtain a uniform bound on the number 
of relevant operators, as long as one makes sure that 
the curvature remains below string scale in order to control the perturbative and 
non-perturbative $\alpha'$-corrections.

\section{The SCHOK Bound}
\label{SCHOKbound}

In a remarkable paper \cite{schh}, Hellerman and Schmidt-Colinet have shown
how modular invariance of the torus partition function can be exploited for 
the purpose of deriving universal bounds on state degeneracies and related 
thermodynamic quantities in 2-dimensional conformal field theories.

Among the many interesting results of \cite{schh} is the statement that a
local conformal field theory of total central charge $c_{\rm tot}=c_L+c_R<48$ and without 
{\it relevant} operators (operators of conformal dimension strictly between $0$ 
and $2$) can have no more than
\begin{equation}
\eqlabel{bound1}
\frac{c_L+c_R}{48-c_L-c_R} \ee^{4\pi} - 2
\end{equation}
{\it marginal} operators (operators of conformal dimension exactly equal to $2$).

The basic idea for deriving \eqref{bound1} is to restrict the modular invariant 
torus partition function $\mathcal{Z}(\tau)$ to purely imaginary $\tau=\ii\beta/(2\pi)$, 
for which $\mathcal{Z}(\tau)$ becomes the thermodynamic partition function 
$\mathcal{Z}(\beta) = \sum_n \ee^{-\beta E_n}$ at temperature $1/\beta$ 
\footnote{Following \cite{schh}, we are assuming here that 
the spectrum of the CFT is discrete, or, in the geometric interpretation, that
the target space is compact.}, and to expand around the self-dual point $\beta=2\pi$. 
To first order, invariance under $\beta\to (2\pi)^2/\beta$ entails a vanishing 
derivative, \ie,
\begin{equation}
\eqlabel{iee}
\left.\frac{d}{d\beta}\right|_{\beta=2\pi} \mathcal{Z}(\beta) = 
\sum_{n} E_n \ee^{-2\pi E_n} = 0 
\end{equation}
One then notes that for sufficiently small central charge, the marginal operators 
contribute states with positive energy in \eqref{iee}. This contribution must be 
balanced by the states of negative energy. Without relevant operators, the only 
state of negative energy is the vacuum. This observation then yields the 
bound \eqref{bound1}. An immediate generalization of this statement that is 
implicit in \cite{schh} is the fact that the number of marginal operators is
bounded from above in terms of the number of operators that are above a certain 
level of relevance. In fact, the above bound is only interesting when 
$c_{\rm tot}\gtrsim 18.27$ for otherwise the CFT necessarily has relevant operators,
as shown in ref.\ \cite{hellerman}. For smaller values of the central charge,
the more general bound still obtains. 

These ideas have been developed in a quantitative way ref.\ \cite{ook}. Keller 
and Ooguri consider 2-dimensional conformal field theories with $\caln=2$ worldsheet 
supersymmetry of central charge $\hat c=3$ and all R-charges in the Neveu-Schwarz
sector integral. The strategy outlined above yields better results after organizing
the contributions into the various representations of the $\caln=2$ superconformal 
algebra extended by the spectral flow (the Odake algebra). As is well-known, 
exactly marginal operators in $\caln=2$ SCFT arise from chiral and antichiral primary 
fields (BPS representations of the $\caln=2$ algebra), and, as shown in \cite{ook},
make a positive 
contribution to the suitably weighted vanishing partition function. When the
$\caln=2$ superconformal field theory describes the IR fixed point of a 
Calabi-Yau sigma model, the number of marginal operators from chiral or twisted 
chiral representations is given by the Hodge numbers $h^{2,1}$ and $h^{1,1}$ of
the Calabi-Yau, respectively; so one uses this notation also for the generic such
SCFT. On the other hand, for central charge $\hat c=3$,
negative contributions to the partition function come only from non-BPS primaries 
of sufficiently small conformal dimension $\Delta_{\rm total}$. Balancing
the positive and negative contributions yields a bound much as in the 
non-supersymmetric case.

The main result of \cite{ook} can be written in the form\footnote{The numbers 
with $\ldots$ are numerical approximations to quantities discussed in \cite{ook}.} 
\begin{equation}
\eqlabel{bound2}
\#\{\text{non-BPS primaries with $\Delta_{\rm total}\le 0.655\ldots$}\}
\ge \frac{1}{522.0\ldots} \bigl(h^{1,1}+h^{2,1}-492.6\ldots\bigr)
\end{equation}
It states that the spectrum of conformal field theories with total Hodge number
$h^{1,1}+h^{2,1}$ sufficiently large must contain non-BPS primary
states of conformal dimension less than $0.655\ldots$, and that
the number of such primary states grows at least linearly in the total
Hodge number. Equivalently, the number of marginal operators is bounded
from above by a linear function of the number of sufficiently relevant operators.

It is worthwhile pointing out that the bound \eqref{bound2} is not necessarily 
optimal. Other variants on the idea of \cite{schh} might yield further refinements
of the basic bound. Our goal is to explore the question whether it is possible
by independent methods to obtain a further upper bound on the number 
of relevant operators, which could then, in combination with the SCHOK bound,
be used to bound the number of marginal deformations in absolute terms.

We note that if any of these sufficiently relevant non-BPS operators survived the 
GSO projection in string theory, they would give rise to tachyonic states in 
space-time. Although they do of course {\it not} survive the GSO projection, we will
therefore refer to the operators on the LHS of \eqref{bound2} as ``counter-factual
tachyons'', or ``tachyons'' for short. For the following discussion
of supplemental geometric bounds, it will be convenient to summarize and remember
the SCHOK bound as the statement that for fixed central charge, there exist 
constants $C_0$, $C_1$ such that in any $\caln=2$ SCFT of that given central charge, 
the number $N_{\rm marginal}$ of exactly marginal BPS operators is bounded linearly 
by the number of tachyons, \ie,
\begin{equation}
\eqlabel{schok}
N_{\rm marginal} < C_0 + C_1 \cdot N_{\rm tachyons}
\end{equation}
Below, we will be mostly concerned with the regime $N_{\rm marginal},
N_{\rm tachyons}\gg 1$. We can then drop $C_0$ from the above statement
without penalty.

\section{Reduction}

The SCHOK bound \eqref{schok} becomes even more interesting when we consider it 
not for isolated conformal field theories, but for the entire family parameterized
by the vevs of the marginal operators. Indeed, while the LHS is constant
in the smooth part of the moduli space of $\caln=2$ SCFT, the RHS is a priori
a strongly moduli dependent quantity. The inequality of course must hold everywhere 
on moduli space, and we can use this freedom to look for regions in the moduli space
in which the number of relevant operators is especially small, or else easy to 
estimate and bound.

Consider in particular an $\caln=2$ SCFT of $\hat c=3$ that can be deformed into 
a phase in which it can be defined by a supersymmetric sigma model into 
a Calabi-Yau threefold $X$.\footnote{We discuss complex dimension $3$ both 
because it is physically the most interesting, and because the SCHOK bound is 
the sharpest in this case. As before, we are assuming that $X$ is compact, and 
the spectrum of the CFT discrete.} Then, the number of marginal operators is 
given by the dimensions of the Dolbeault cohomology groups $H^{1,1}(X)$, 
parameterizing K\"ahler deformations, and  $H^{2,1}(X)$, parameterizing complex 
structure deformations. Together, $H^{1,1}(X)\oplus H^{2,1}(X)$ is the tangent 
space to the space of Ricci-flat K\"ahler metrics on $X$ (Yau's theorem).

On the other hand, to leading order in the $\alpha'$-expansion of the string worldsheet 
theory (this is, morally speaking, the ``supergravity approximation''), 
any would-be tachyonic states must be understood to arise by ``Kaluza-Klein 
reduction'' of the center-of-mass motion of the string with at most one 
$\psi^{\mu}_{-1/2}$ oscillator excited. In first approximation, the conformal 
dimensions of primaries, $\Delta_n$, are given by the eigenvalues, $\lambda_n$, 
of the Laplacian acting on scalar or vector-valued wavefunctions on $X$,
\begin{equation}
\eqlabel{roughly}
\Delta_n = \alpha' \lambda_n + \cdots
\end{equation}
In the limit in moduli space in which the volume of $X$ becomes very large, the 
eigenvalues accumulate at zero, so that the bound \eqref{bound2} will
be trivially satisfied.\footnote{Here, we are anticipating Weyl's law, to be 
discussed further below.} Indeed, the LHS is just the dimension
of the moduli space, and of course constant in the limit.

Conversely, the bound becomes potentially stringent if we can deform the manifold
to a region in the moduli space of Ricci-flat metrics in which the number of 
low-lying eigenvalues of the Laplace operator is small. Intuitively, this will 
happen for a manifold of small volume. Unfortunately, in the small volume limit, 
the supergravity approximation will break down, $\alpha'$-corrections will become 
large, and we will not be able to estimate the number of tachyons by counting 
eigenvalues of geometric operators. (By flat space intuition, the light spectrum
will be dominated by winding modes in this regime \cite{doga}.)

The strategy we advocate is to look for an intermediate regime in which the 
supergravity approximation is valid (say, the curvature radius and volume of the 
manifold are large 
in string units) yet the manifold is not so large that the continuum of states 
has fully materialized. Of course, a bound on the number of relevant operators 
in the CFT that one might obtain in this regime will be far from optimal. On 
the other hand, assuming \eqref{roughly} reduces the problem to a question amenable to
exact mathematical analysis, which we discuss in the following sections.
The problem in this regime remains non-trivial and highlights what we believe
are the essential challenges in bounding the number of relevant operators more generally.

The replacement of the 2-d sigma model by its point-particle approximation itself
is difficult to justify rigorously. There exists substantial evidence for the
conjecture that a Calabi-Yau sigma model flows to an $\caln=2$ SCFT in the
infrared that admits an independent, and mathematically rigorous, construction in
certain cases. This evidence is based on a comparison of the massless spectrum,
\ie, the cohomology of the manifold, which is BPS and therefore protected by
supersymmetry against renormalization, and on the general structure of 
higher-order terms in the $\beta$-function of $\caln=2$ sigma models.
For the non-BPS spectrum however, the perturbative $\alpha'$-expansion 
\eqref{roughly} is not expected to be better than standard asymptotic 
expansions in quantum field theory. On the other hand, and this is an important
distinction to the discussion in the following sections, the corrections
are local in target space, and expected to be uniformly suppressed under 
the assumption that the curvature radius is large in string units.
Related issue in a similar regime, albeit with somewhat 
different aims, were discussed recently in \cite{doga}.

\section{On Uniform Bounds}

Given \eqref{schok}, in order to bound the dimension of moduli space, 
$N_{\rm marginal}$, of $\caln=2$ SCFTs, it is enough to find, in the moduli
space of deformations of any given SCFT, a point or region in which $N_{\rm
tachyons}$ is bounded by some universal constant. Let us recall that 
$N_{\rm tachyons}$ is defined as the number of primary states with conformal 
dimension below $0.655\ldots$, see \eqref{bound2}.

As explained in the previous section, we will restrict to those $\caln=2$
SCFTs which have a Calabi-Yau phase in their moduli space, and we will look
for the relevant region in the vicinity of the large-volume regime. 
We will assume 
that the conformal dimensions of relevant operators in the CFT can be approximated 
by the low-lying eigenvalues of the Laplace operator of the Ricci-flat metric 
on the Calabi-Yau, eq. \eqref{roughly}, given that the curvature radius is 
large in string units. We will for simplicity restrict to scalar wavefunctions. 
We expect vector-valued wavefunctions to behave qualitatively similarly as long 
as the manifold is simply connected. Forms of higher degree and other geometric 
operators are not expected to play a role since the corresponding modes are 
already massive in the flat space limit.

For the geometric analysis, it is sometimes convenient to study instead of the 
distribution of eigenvalues itself, its Laplace transform,
\begin{equation}
\eqlabel{heattrace}
Z_X(t) = \sum_{n} \ee^{-t\lambda_n} = \tr\ee^{-t\Delta}
\end{equation}
which is otherwise known as the trace of the heat kernel in the literature. We 
may intuitively identify it as the point-particle approximation to the full 
stringy partition function. Here, 
\begin{equation}
\Delta = -\frac{1}{\sqrt g}\frac{\partial}{\partial x^I} g^{IJ} \sqrt{g}
\frac{\partial}{\partial x^J}
\end{equation}
is the (positive-definite) scalar Laplacian on $X$, with metric $g_{IJ}$.

To reformulate the bound in terms of $Z_X(t)$, we note that for any fixed $t_*$,
$Z_X(t_*)$ is bounded below by $\ee^{-1}$ times the number of eigenvalues smaller
than $1/t_*$. Therefore, for $t_*=\alpha'/0.655\ldots$, an upper bound on $Z_X(t_*)$
implies an upper bound on $N_{\rm tachyons}$. In the regime of concern, we may
then write the SCHOK bound \eqref{schok} as
\begin{equation}
\eqlabel{schokp}
N_{\rm marginal} < \ee\cdot C_1 \cdot Z_X(t_*).
\end{equation}
Note that we are assuming that $X$ is large and weakly curved in string units, but that we 
are otherwise allowing arbitrary (K\"ahler and complex structure) deformations of the metric in 
order to make $Z_X(t_*)$ as small as possible.

Then, as an approximation to an effective bound on the number of tachyons, we may 
ask the following mathematically sharp question: \\[.2cm]
\framebox{\parbox{\textwidth-.3cm}{\it Does there exist a constant $B$ such
that for any diffeomorphism class of Calabi-Yau manifolds (of fixed dimension $3$),
there exists a Ricci-flat K\"ahler metric on a representative $X$ with sufficiently 
large volume and curvature radius in string units, such that $Z_X(t_*)<B$?}}
\\[.2cm]
The bound is uniform in the topology of the manifold, in the sense that the constant
$B$ is universal. The curvature radius should be (also uniformly) large in string units 
in order to be able to control the $\alpha'$ corrections as explained above.

We have formulated this question in terms of the behaviour of $Z_X(t)$ for fixed $t=t_*$ 
and varying metric on $X$. However, the mathematical results more readily control $Z_X(t)$ 
for fixed $X$ and varying $t$. To fix ideas, let us recall here the most well-know result 
on eigenvalue asymptotics, Weyl's law. It states that for fixed $X$, asymptotically as 
$t\to 0$,
\begin{equation}
\eqlabel{weyl}
Z_X(t) \sim (4\pi t)^{-\dim_{\mathbb{R}}(X)/2}\vol(X)+\ldots
\end{equation}
We note right away that since \eqref{weyl} is only an asymptotic expansion for
given $X$, it can not be used to bound $Z_X(t)$ directly by merely controlling the
overall volume of $X$. Indeed, as we vary $X$, the expansion might be valid only for
ever smaller values of $t$.

In order to facilitate the mathematical analysis, we will make one more reformulation.
The eigenvalues of the Laplace operator scale as $R^{-2}$ under 
overall rescaling of the metric on $X$ by a factor of $R$. Therefore, $Z_X(t)$ depends 
only on the combination $t/R^2$. To make this explicit, it is convenient to 
separate the overall scale of 
the metric on $X$ from the remaining deformations. The moduli space $\calm(X)$
of Ricci-flat K\"ahler metrics on a Calabi-Yau manifold is finite-dimensional
and locally the product of the complex structure deformations and the 
K\"ahler cone. We will write it as
\begin{equation}
\calm(X) = \calm_1(X) \times \reals_+
\end{equation}
where $\calm_1\ni m$ is the moduli space of manifolds of unit volume (in string units),
and $\reals_+\ni R$ parameterizes the overall scale. We will refer to the
manifold with fixed moduli by $X_{m,R}$. 

We can then reformulate the above question as follows: \\[.2cm]
\framebox{\parbox{\textwidth-.3cm}{
\textit{Does there exist 
a constant $B$ and a scale $R_*\gg 1$ such that every diffeomorphism class of 
Calabi-Yau manifolds admits a metric representative $X_{m,R}$ with $R\approx R_*$ 
and radius of curvature of $X_{m,1}$ no smaller than $(\sqrt{\alpha'}R_*)^{-1}$ such that
\begin{equation}
\eqlabel{question}
Z_{X_{m,R}}(t_*) = Z_{X_{m,1}}(t_*/R^2) < B ?
\end{equation}}}}
\\[.2cm]
In the remaining parts of this paper, we will write the LHS simply as $Z_X(t_*)$, 
with the understanding that $X$ has volume 1 and that the (large) $R$-modulus is 
absorbed into $t_*$. We thus have the freedom to vary $t_*$ away from
$\alpha'/0.655\ldots$, but it should be remembered that it will ultimately be a fixed 
small parameter.

\section{Large Volume Expansion and Curvature Singularities}

The basic intuition about question \eqref{question} comes from Weyl's law, which 
depends {\it only} on the volume of the manifold. To better 
understand which conditions could then possibly {\it prevent} the existence of 
a universal bound, we will begin by examining the higher-order terms in the asymptotic 
expansion. We will eventually find them to be insufficiently precise for our purposes. 
However, as we shall 
discuss momentarily, they exhibit the critical role played by curvature singularities 
pertaining to our question.

\subsection{Heat trace asymptotics}

Weyl's law is the first term of a well-known series expansion in spectral geometry 
building up on pioneering work by Minakshisundaram and Pleijel. We consider the
general heat trace $\tr_X f e^{-t\Delta}$ on a smooth Riemannian 
manifold $X$. Here $f\in C^\infty(X)$ is an optional function used 
for localization \cite{gilkey}. For $f=1$, $\tr_X f e^{-t\Delta}$ reduces to $Z_X(t)$. 
The general heat trace admits the following asymptotic 
expansion \cite{gilkey,vassilevich}
\begin{equation}
\eqlabel{expansion}
\tr_X f e^{-t\Delta}\sim\frac{1}{t^d}\sum_{n=0}^\infty 
[a^X_n(f)+a^{\partial X}_n(f)]t^{n/2},\qquad t\to 0,
\end{equation}
where $d=\dim_{\mathbb{R}}(X)$, $a^X_{2k+1}(f)=0$, and $a^X_{2k}(f)$ are curvature 
invariants of degree $k$ of $X$. The first few read
\begin{align}
\eqlabel{a0}
a_0^X(f)&=\frac{1}{(4\pi)^{d/2}}\int_X f \sqrt{g}d^dx \\
\eqlabel{a2}
a_2^X(f)&=\frac{1}{(4\pi)^{d/2}}\frac{1}{6}\int_X Rf \sqrt{g}d^dx \\
\eqlabel{a4}
a_4^X(f)&=\frac{1}{(4\pi)^{d/2}}\frac{1}{360}\int_X [12\Delta R+5R^2-2R_{ij}R^{ij}+
2R_{ijkl}R^{ijkl}]f \sqrt{g}d^dx \\
a_6^X(f)&=\frac{1}{(4\pi)^{d/2}}\frac{1}{5040}\int_X [18\Delta^2R+17\nabla^kR\nabla_kR
-2\nabla^kR_{ij}\nabla_kR^{ij}-4\nabla_nR_{jk}\nabla^kR^{jn} \nonumber \\
&+9\nabla^nR_{ijkl}\nabla_nR^{ijkl}+28R\Delta R-8R_{jk}\Delta R^{jk}+24R_{jk}\nabla^k
\nabla_nR^{jn}+12R_{ijkl}\Delta R^{ijkl} \nonumber \\
&+\frac{35}{9}R^3-\frac{14}{3}RR_{ijkl}R^{ijkl}-\frac{208}{9}R_{jk}{R^j}_nR^{kn}+
\frac{64}{3}R_{ij}R_{kl}R^{ikjl}-\frac{16}{3}{R^j}_kR_{jnli}R^{knli} \nonumber \\
&+\frac{44}{9}{R^{ij}}_{kn}R_{ijlp}R^{knlp}+\frac{80}{9}R^{ijkn}R_{ilkp}{{{R_j}^l}_n}^p]f 
\sqrt{g}d^dx. \eqlabel{a6}
\end{align}

As expected from Weyl's law, $a_0^{X}(1)$ reduces to the volume of $X$. Some higher order 
coefficients are readily available in the literature, but they quickly 
become non-manageable for practical purposes. Following usual conventions, $R_{ijkl}$, 
$R_{ij}$, and $R$ (not to be confused with the volume modulus introduced above) are the 
Riemann, Ricci, and scalar curvatures, respectively. 
Indices are contracted with the metric $g$ and 
covariant derivatives are in the Levi-Civita connection. Note that for a Calabi-Yau
manifold with Ricci flat metric, $a_2=0$ and the higher-order terms simplify significantly
as well.

The terms $a^{\partial X}_k(f)=
\int_{\partial X}...$ are integrated boundary invariants involving the curvature and the 
embedding of $\partial X\subset X$. Their precise expression depends on whether Dirichlet, 
Neumann, or mixed conditions are imposed at the boundary. For Dirichlet boundary 
conditions, the leading of these are \cite{gilkey}
\begin{align}
a_0^{\partial X}&=0 \\
a_1^{\partial X}&=-\frac{1}{(4\pi)^{(d-1)/2}}\frac{1}{4}\int_{\partial X}f\sqrt{h}d^{d-1}y.
\end{align}
We are using $h$ as the induced metric on $\partial X$.

In order to use the expansion \eqref{expansion} for the purpose of {\it bounding}
$Z_X(t)$ as required in \eqref{question}, we would need to estimate the remainder
after truncation to some finite order. We are not aware of such 
estimates. But if, instead, we look at a possible bound on any given term (which 
would be enough if the series were convergent), it appears that the central 
problem is to control the behaviour of the heat trace in regions of high curvature 
(as compared to the volume, but low as compared to the string 
scale)\footnote{At first, it seems surprising that controlling the number of 
tachyons, which sounds like an infrared property of the manifold, should involve 
curvature singularities, which are visible by probing short distances. However, the
reformulation \eqref{question} makes it clear that it is indeed 
the UV properties of the manifold of unit volume $X_{m,1}$ that determine
the long-distance spectrum of the manifolds we are ultimately after.}.
In particular, a uniform bound could fail to exist if, under exploration of the 
large volume region in the moduli space (which we recall is the basic premise of our
strategy), curvature singularities with uncontrolled
contribution to the heat trace were unavoidable.

The conifold singularity \cite{conifold} is both the proto-typical example of a 
curvature
singularity of Calabi-Yau manifolds, and arises generically under deformation
of the manifold. Certainly all standard constructions of Calabi-Yau manifolds 
known to us unavoidably lead to conifold singularities somewhere in their 
moduli space. Therefore, studying possible bounds on the trace of the heat
kernel under the approach to the conifold singularity (from either the deformed
or the resolved side) is a natural first test case for answering question 
\eqref{question}.

\subsection{Conifolds as local models for curvature}

Since the resolved conifold is more symmetric\footnote{The small resolution of the conifold 
singularity breaks only a $\mathbb{Z}_2$ in the $O(4)\times U(1)$ symmetry of the 
singular conifold, while the deformation breaks a whole $U(1)$ to a $\mathbb{Z}_2$.}, 
we exclusively treat it in our detailed analysis, with the expectation that other 
approaches to Calabi-Yau singularities are not qualitatively different. We will 
discuss the lessons that we learn for the general case in our concluding section. 
Certainly, the patching 
procedure, that we are about to explain, is simple enough that it carries readily 
to any model of curvature that one might want to examine. A further advantage of
the conifold singularity is that the moduli problem becomes effectively
one-dimensional.

Let us then assume that the Calabi-Yau metric on the compact boundary-free $X$ can be 
approached on some open $U\subset X$ by the resolved conifold metric 
\cite{pandozayas,klebanovResolved}
\begin{equation}
\eqlabel{resolvedmetric}
ds^2_{\hat{C}}=\frac{r^2+6\epsilon^2}{r^2+9\epsilon^2}dr^2+\frac{r^2}{9}
\frac{r^2+9\epsilon^2}{r^2+6\epsilon^2}\Big(d\psi+\sum_{k=1}^2 \cos
\theta_k d\phi_k\Big)^2+\frac{r^2}{6}ds^2_{S^2_{(1)}}+\frac{r^2+6\epsilon^2}{6}
ds^2_{S^2_{(2)}},
\end{equation}
\begin{equation}
r\in\mathbb{R}^+,\quad \psi\in[0,4\pi),\quad \theta_k\in[0,\pi),\quad \phi_k\in[0,2\pi)
\end{equation}
The round metric on spheres are parameterized as $ds^2_{S^2_{(k)}}=d\theta_k^2+
\sin^2\theta_k d\phi_k^2$. The only dimensionful coordinate, $r$, measures the 
distance in string units from the 
$S^2$ (of radius $\epsilon$) at the bottom of the geometry. At some $r_\infty\gg 
\epsilon$, the resolved conifold model effectively ceases to be valid, but the 
resolved metric is nevertheless assumed to interpolate smoothly to the unknown metric 
on $X\backslash U$.

\begin{figure}[bht]
\begin{center}
\includegraphics[scale=1]{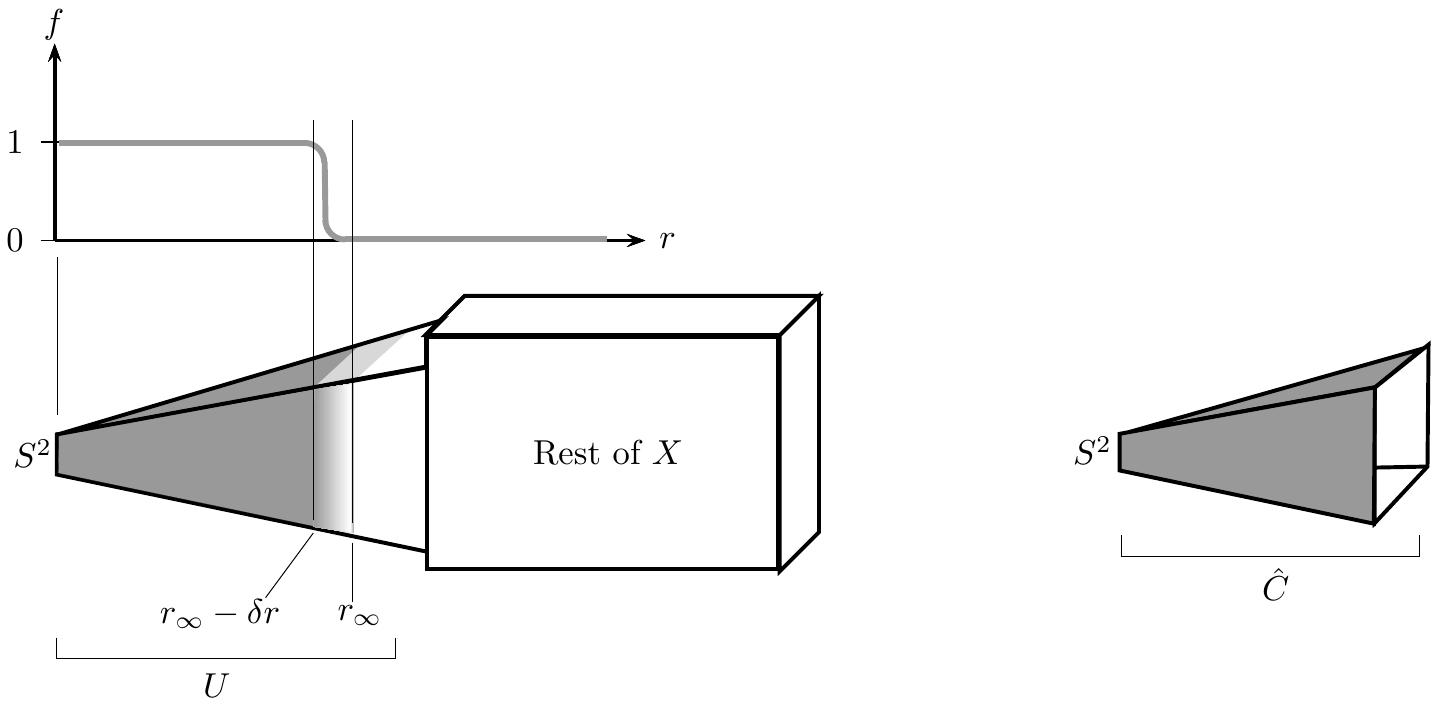}
\end{center}
\caption{(left) Localization function on $X$ allowing a local study of 
the heat trace asymptotics; (right) Sharp cut-off at finite radius.}
\label{fig1}
\end{figure}

As our metric is only explicit on $U\subset X$, it is convenient to require the accessory 
function $f$ to vanish on $X\backslash U$ and, on $U$, to equal $1$ for 
$r<r_\infty-\delta r$ ($\delta r \ll r_\infty$), to vanish for $r>r_\infty$, and to 
vary smoothly from $1$ to $0$ on the interval $[r_\infty-\delta r,r_\infty]$ 
(see fig.\ \ref{fig1}, left). 
With this choice, we have effectively excised the resolved conifold-like patch $\hat{C}$ 
from the compact threefold while maintaining the boundary contributions 
$a^{\partial X}_n(f)$ equal to $0$. On $X$, these
terms would vanish for any $f\in C^\infty(X)$ because the space is 
boundary-free, while on $\hat{C}$ they vanish because our chosen $f$ is zero on 
$\partial\hat{C}$.

In practice, we however avoid dealing with a ``fuzzy boundary'' by 
declaring $\hat{C}$ 
to be the resolved conifold \textit{sharply} truncated at $r=r_\infty$ 
(fig.\ \ref{fig1}, right). 
From now on, we deal with the heat trace $Z_{\hat{C}}(t,\epsilon)=
\tr_{\hat{C}} e^{-t\Delta}$ on this space. Following \eqref{question}, we address 
the prospect of \textit{bounding $Z_{\hat{C}}(t_*,\epsilon)$ independently of 
$\epsilon$ in the limit $\epsilon\to 0$ for $t_*$ fixed but small}.

\subsection{Asymptotics on the resolved conifold}

A side-effect of the cut-off is to generate spurious heat trace boundary contributions, 
which are ultimately of no interest. Assuming Dirichlet boundary conditions (a choice 
to which we will stick all along), the leading of these terms is a negative contribution:
\begin{equation}
a_1^{\partial \hat{C}}(1)=-\frac{\sqrt{\pi}}{216}r_\infty^5
\sqrt{\left(1+\frac{6\epsilon^2}{r_\infty^2}\right)\left(1+
\frac{9\epsilon^2}{r_\infty^2}\right)}.
\end{equation}

However, the now well-posed geometry (sharply truncated) only differs superficially 
from the smoother one excised from $X$, so we expect bulk spectral behaviour to be the 
same. Except now, \eqref{a0}--\eqref{a6} can be conveniently used to calculate 
explicitly $a_{2n}^{\hat{C}}(1)$, $n=0,1,\ldots$ These are the bulk contributions 
attributable to the resolved conifold-like patch in the heat trace expansion on 
(the more untractable space) $X$. We have calculated them exactly up to order $6$:
\begin{align}
\eqlabel{a0calc}
a_0^{\hat{C}}(1)&=\frac{r_\infty^6}{648}\left(1+\frac{9\epsilon^2}{r_\infty^2}\right) \\
\eqlabel{a2calc}
a_2^{\hat{C}}(1)&=0 \\
\eqlabel{a4calc}
a_4^{\hat{C}}(1)&=\frac{r_\infty^2}{810}\left(1+\frac{6\epsilon^2}{r_\infty^2}
\right)^{-4}\left(2+\frac{45\epsilon^2}{r_\infty^2}+\frac{360\epsilon^4}{r_\infty^4}
+\frac{1080\epsilon^6}{r_\infty^6}\right) \\
\notag
a_6^{\hat{C}}(1)&=\frac{1}{8505}\ln\left(1+\frac{r_\infty^2}{6\epsilon^2}\right)-
\frac{1}{99225}\left(1+\frac{6\epsilon^2}{r_\infty^2}\right)^{-7}\left(284+
\frac{11963\epsilon^2}{r_\infty^2}+\frac{214564\epsilon^4}{r_\infty^4}\right.
\\
&\qquad
\eqlabel{a6calc}
\left.+\frac{2127300\epsilon^6}{r_\infty^6}+\frac{12594960
\epsilon^8}{r_\infty^8}+\frac{35789040\epsilon^{10}}{r_\infty^{10}}+
\frac{544320\epsilon^{12}}{\epsilon^{12}}\right)
\end{align}

The dominant term $a_0^{\hat{C}}(1)$ is proportional to $\vol\hat{C}$; a consequence of 
Weyl's law \eqref{weyl} applied here to $\hat{C}$ rather than $X$. The effect of the 
parameter $\hat{R}=(\vol\hat{C})^{1/6}$ was previously addressed above, so we might want to 
fix it in order to focus on the intrinsic effect of changing the radius of the $S^2$. 
More conveniently, we will allow it to vary continuously as $\epsilon\to 0$ and 
absorb the change of volume in $X\backslash U$ in such a way that $\vol X=1$ is 
constant all the way in the limit. The sub-leading term $a_2^{\hat{C}}(1)$ vanishes 
by virtue of Ricci-flatness, so the first non-trivial signature of finite $\epsilon$ 
emerges at order 4. It is noteworthy that $a_4^{\hat{C}}(1)$ is positive and bounded 
in the limit $\epsilon\to 0$.

The next term $a_6^{\hat{C}}(1)$ however starts exhibiting logarithmic \textit{divergence}. 
A simple argument suggests the leading divergence as $\epsilon\to 0$ at higher orders.
On the vanishing $S^2$ at $r=0$, the curvature behaves as $R\sim 1/\epsilon^2$. The 
curvature enters $a_{2k}^{\hat{C}}(1)$ through its $k$-th power, thus yielding a 
$\sim\epsilon^{-2k}$ behaviour. Finally, the volume integral contributes an extra 
shift of the power by the dimension:
\begin{equation}
\eqlabel{ablowup}
a_{2k}^{\hat{C}}(1)\sim \epsilon^{6-2k},\qquad(k>3).
\end{equation}
These remarks seemingly imply that, no matter how small we make $t_*$, as 
$\epsilon$ approaches $0$, $Z_{\hat{C}}(t_*,\epsilon)$ will grow without bounds. 
If this were the case, we would at the least need to stay clear of any conifold 
singularities in the moduli space in order to obtain a bound of the type 
\eqref{question}. For a single localized singularity such as on $\hat C$, 
this can be achieved by simply making the resolution parameter large enough.
However, in the presence of numerous shrinkable 2- and 3-cycles in the complete
manifold $X$, avoiding the formation of all the potential singularities could
prevent the bound from being uniform in the topology of $X$.

\begin{figure}[h]
\begin{center}
\includegraphics[angle=90,origin=c]{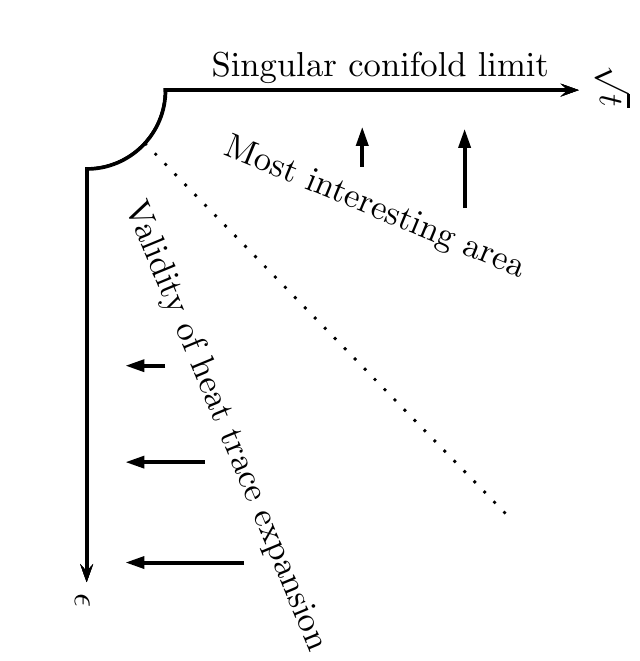}
\end{center}
\caption{Domain of $Z_{\hat{C}}(t,\epsilon)$ blown up at the origin}
\label{fig2}
\end{figure}

On the other hand, the expansion itself is only guaranteed to be valid for fixed 
geometry and $t\to 0$. The question we have asked instead concerns the limit 
$\epsilon\to 0$ for small (but fixed) $t$. It is possible that 
$Z_{\hat{C}}(t,\epsilon)$, thought of as a function $\mathbb{R}_{\epsilon}^+
\times\mathbb{R}_{\sqrt{t}}^+\to\mathbb{R}^+$, does not admit a \textit{joint} 
expansion in ($\epsilon$,$\sqrt{t}$), but that it is still bounded despite
what the expansion in one of the variables might lead us to suppose. 
Indeed, the general mathematical expectation (see for instance, \cite{mazzeo})
is that the trace of the heat 
kernel is well-behaved on the \textit{(real) blow up} of $\mathbb{R}_{\epsilon}^+
\times\mathbb{R}_{\sqrt{t}}^+$ (fig.\ \ref{fig2}) at the origin.\footnote{A useful 
toy example of this kind of behaviour is provided 
by the function $\arctan({\sqrt{t}}/{\epsilon})$. For fix $\epsilon$, 
$\arctan({\sqrt{t}}/{\epsilon})\sim{\sqrt{t}}/{\epsilon}-{t^{3/2}}/{3
\epsilon^3}+O\left(\left({\sqrt{t}}/{\epsilon}\right)^5\right)$, as 
$\sqrt{t}\to 0$. All of these terms are unbounded when $\epsilon\to 0$ even 
though $\arctan({\sqrt{t}}/{\epsilon})$ is bounded above by $\pi/2$ on the blow 
up space (which basically amounts to using radial coordinates on $(\epsilon,
\sqrt{t}$-space
in this case).} Proving such claims 
however involves dwelling into the realm of microlocal analysis (see in particular 
Melrose \cite{melrose1,melrose2}). 
An alternative route, described in the next section, is to take advantage of our 
explicit metric to examine the spectral properties of $\hat{C}$. This approach 
already provides compelling evidence for the existence of a bound, as we shall 
see.

\subsection{Asymptotics on spaces with conical singularities}

Let us close this section by pointing out that the analysis of the heat trace can 
be carried out as well on the singular conifold $C$ (\ie, in the limit $\epsilon=0$ 
itself). The question of whether the heat trace displays any kind of divergence as 
$\epsilon\to 0$ can then be reformulated as a question of \textit{continuity} 
in the same limit.

The heat trace asymptotics on spaces with conical singularities was studied 
by Cheeger \cite{cheeger}. The first step of the analysis is to ensure that
the eigenvalue problem is well-defined on the singular space. For the 
scalar problem, this can easily be verified by separation of variables and
reduction to a one-dimensional problem, which we will study in the next
section. Quite similarly, one can analyze the Laplacian on forms of arbitrary
degree on a space with conical singularities. One finds that this operator
is essentially self-adjoint (so the eigenvalue problem is well-defined by
itself) unless the base of the cone has non-trivial middle-dimensional cohomology
(in which case a choice of boundary condition at the singularity is required to
make the problem well-defined). 

Cheeger then showed that the heat trace (for forms of arbitrary degree) 
on a manifold with conical singularities
admits an asymptotic expansion very similar in essence to \eqref{expansion} 
(see also \cite{kirsten} for an application to metric cones). Some 
of the terms of this expansion can be thought of as the above \eqref{a0}--\eqref{a6} 
having been ``regulated'' in order to tame the contribution from the infinite curvature 
at the tip. More precisely, a $u$-sized conical patch is removed from the 
manifold (much like we did in subsection 5.2) and the integrals in \eqref{a0}--\eqref{a6} are 
taken over the remaining space. In the limit $u\to 0$, the integrals do not converge 
but the infinite term can be unambiguously identified and subtracted. The finite 
part that is left is what enters the asymptotic expansion on the singular manifold. It 
is noteworthy that the infinite parts diverge as $\ln(u)$ (for $n=d$, where $d$ is 
the real dimension of the manifold) and $u^{d-k}$ (for $n>d$). This is strongly 
reminiscent of our situation (\cf~\eqref{a6calc},\eqref{ablowup}).

We have not tried to make more precise the intuition that the finite parts 
in \eqref{a0calc}--\eqref{a6calc} are the relevant ones for the heat trace
expansion in the singular limit $\epsilon=0$. However, the existence of
the expansion at $\epsilon=0$ together with the general expectation mentioned
above, is reasonable evidence that the divergence of the coefficients is merely 
an artifact of the asymptotic expansion for $t\to 0$ at fixed $\epsilon$,
whereas the full heat trace can still be continuous in the limit $\epsilon \to 0$ for
finite $t$. If this is the case, the asymptotic expansion on the singular 
manifold would be a useful starting point for estimating $Z_{\hat C}(t,\epsilon)$
on the resolution. With a resolution of order $1/R_*$, this estimate could
be useful to establish a bound of the form \eqref{question} on $Z_X(t)$ for
the complete manifold $X$.

An alternative argument against the persistence of the divergences goes as follows. 
In the degenerate case $\epsilon=0$, additional contributions to the $t$-independent 
terms and extra $\sim\ln t$ terms need to be added to the heat trace expansion 
\eqref{expansion} \cite{cheeger}. There is thus a \textit{qualitative discontinuity 
in the nature of the two expansions}. The infinities might thus reflect the fact that 
the expansion ceases to be a valid representation of $Z_{\hat{C}}(t,\epsilon)$, rather 
than indicating a divergence in the heat trace itself. If the transition is indeed 
continuous, the infinities coming from the expansion at finite $\epsilon$ must 
conspire (through some re-summation of the divergent asymptotic series) to yield 
the logarithms and additional terms in the expansion for $\epsilon=0$.

\section{Exact Spectral Analysis on Conifolds}

Given the limitations of the asymptotic approach, we now consider more closely the 
spectrum of the scalar Laplacian on $\hat{C}$. The question of boundedness of 
$Z_{\hat{C}}(t,\epsilon)$ in the limit $\epsilon\to 0$ can be recast in a question 
about the behaviour of the eigenvalues in the same limit. Let us denote 
$N_{\hat{C}}(\lambda,\epsilon)$ the number of eigenvalues of $\Delta$ smaller 
than $\lambda$. We will call it the (full) \textit{counting function} of the 
eigenvalues. Suppose it is bounded above (for $\epsilon$ small enough) by some 
function $B(\lambda)$ independent of $\epsilon$. Then the fact that 
$Z_{\hat{C}}(t,\epsilon)$ is the Laplace transform of the derivative of 
$N_{\hat{C}}(\lambda,\epsilon)$ with respect to $\lambda$ entails readily an 
upper bound on the heat trace itself.\footnote{A bound also arise if the counting 
function is bounded only beyond some ($\epsilon$-independent) $\lambda_*$.} In 
terms of the individual eigenvalues, bounding the counting function is equivalent 
to having only finitely many eigenvalues dropping below any fixed (large enough) 
$\lambda$ as $\epsilon\to 0$.

In this section, we provide an analytical analysis of the eigenvalue problem and 
discuss its ramifications and limitations. In section 7, we come back to it using 
high-order WKB expansions. This approach and the above formulation in terms of 
$N_{\hat{C}}(\lambda,\epsilon)$ allow us to finally claim a \textit{positive} 
answer to the question raised at the end of section~5.2.

\subsection{Singular conifold}

An advantage of considering first the singular conifold is that the eigenvalue 
problem of the scalar Laplacian on this space is completely amenable to 
separation of variables. Its solution, which we now briefly review, offers a 
benchmark to refer to when analyzing spaces resolving the singularity.

Upon setting $\epsilon=0$ in \eqref{resolvedmetric}, $\hat{C}$ becomes 
manifestly the metric cone $C$ over $T^{1,1}$ (still truncated at a finite 
$r=r_\infty$). For cones, the eigenvalue problem splits into a ``radial'' differential 
equation, which determines the spectrum, and an eigenvalue problem on the base 
space. Letting the eigenfunctions be denoted
\begin{equation}
f^{n,m,l_1,m_1,l_2,m_2}(r,\psi,
\theta_1,\phi_1,\theta_2,\phi_2)=\text{R}^{n,m,l_1,l_2}(r)\Psi^m(\psi)
\prod_{k=1}^2\Theta_k^{m,l_k,m_k}(\theta_k)\Phi_k^{m_k}(\phi_k),
\end{equation}
the radial equation,
\begin{equation}
\eqlabel{radialeq0}
\frac{1}{r^5}\frac{d}{dr}\left(r^5\frac{d}{dr}\text{R}^{m,l_1,l_2}(r)\right)+
\left(\lambda-\frac{\Lambda^{m,l_1,l_2}}{r^2}\right)\text{R}^{m,l_1,l_2}(r)=0,
\end{equation}
is Bessel differential equation (up to the change of variable $r\to\sqrt{\lambda}r$ 
($\lambda\neq 0$) and factorizing $(\sqrt{\lambda}r)^{-2}$ out of 
$\text{R}^{m,l_1,l_2}(r)$). Requiring the eigenfunctions to be regular on 
$[0,r_\infty]$ and Dirichlet boundary condition at $r=r_\infty$ picks out 
the solution of the first kind
\begin{equation}
\text{R}^{m,l_1,l_2}(r)\propto\frac{1}{\lambda r^2} 
J_{\sqrt{\Lambda^{m,l_1,l_2}+4}}(\sqrt{\lambda}r).
\end{equation}
The spectrum is determined 
from the strictly positive zeros of the Bessel function:
\begin{equation}
J_{\sqrt{\Lambda^{m,l_1,l_2}+4}}(\sqrt{\lambda}r_\infty)=0.
\end{equation}

Here $\Lambda^{m,l_1,l_2}$ are Laplacian eigenvalues on the base manifold $T^{1,1}$. 
Along with the eigenfunctions, they have been worked out in 
\cite{klebanovResolved,gubser,ceresole}:
\begin{equation}
\eqlabel{basespectrum1}
\Lambda^{m,l_1,l_2}=9m^2+\sum_{k=1}^2\Lambda_k^{m,l_k},\quad \Lambda_k^{m,l_k}=6[l_k(l_k+1)-m^2],
\end{equation}
\begin{equation}
\eqlabel{basespectrum2}
\Psi^m(\psi)=e^{im},\quad m\in\{-\min(l_1,l_2),...,+\min(l_1,l_2)\} \text{ (integer increments)}
\end{equation}
\begin{equation}
\eqlabel{basespectrum3}
\Phi_k^{m_k}(\phi_k)=e^{im_k},\quad m_k\in\{-l_k,...,+l_k\} \text{ (integer increments)}
\end{equation}
\begin{equation}
\eqlabel{basespectrum4}
\Theta_k^{m,l_k,m_k}(\theta_k)=\left\{
\begin{array}{l l}      
    \sin^{m_k}\theta_k\cot^{m}\frac{\theta_k}{2}{}_2F_1({\scriptstyle -l+m_k},{\scriptstyle 1+l+m_k},{\scriptstyle 1+m_k-m_{~}},
\sin^2\frac{\theta_k}{2}) & m_k\geq m\\
    \sin^{m}\theta_k\cot^{m_k}\frac{\theta_k}{2}{}_2F_1({\scriptstyle -l+m_{~}},{\scriptstyle 1+l+m_{~}},{\scriptstyle 1+m_{~}-m_k},
\sin^2\frac{\theta_k}{2}) & m_k\leq m
\end{array}\right.
\end{equation}
with either both $l_k\in\{0,1,2,...\}$ or both $l_k\in\{\frac{1}{2},\frac{3}{2},
\frac{5}{2},...\}$, $k=1,2$.

Note that harmonic functions, corresponding to $\lambda=0$, are irrelevant here 
since the only solutions to $\Delta f=0$ (being thought of as the steady state 
source-free heat equation) are constant functions over $C$. The Dirichlet 
condition at the cut-off then picks up $f=0$, which has to be discarded.

\subsection{Resolved conifold}

Since the resolved and singular conifolds share the same $SU(2)_{l_1,m_1}
\times SU(2)_{l_2,m_2}\times U(1)_m$ symmetry (at the level of the algebra), 
the eigenvalue problem on $\hat{C}$ 
is structurally identical to the above. Separation of variables yields again a 
complete solution and, in fact, the ``angular'' parts of the eigenfunctions 
remain unchanged (\ie, \eqref{basespectrum1}--\eqref{basespectrum4} are still 
valid). Moreover, the spectrum information is still encoded in a radial ordinary 
differential equation. It takes the form
\begin{equation}
\eqlabel{radialeq1}
\frac{1}{r^3(r^2+6\epsilon^2)}\frac{d}{dr}\left(r^3(r^2+9\epsilon^2)
\frac{d}{dr}\text{R}^{m,l_1,l_2}(r,\epsilon)\right)+\left(\lambda-
\frac{\Lambda^{m,l_1,l_2}(r,\epsilon)}{r^2}\right)
\text{R}^{m,l_1,l_2}(r,\epsilon)=0
\end{equation}
where
\begin{equation}
\frac{\Lambda^{m,l_1,l_2}(r,\epsilon)}{r^2}=\frac{\Lambda_1^{m,l_1}}{r^2}+
\frac{\Lambda_2^{m,l_2}}{r^2+6\epsilon^2}+
\frac{9m^2}{r^2}\frac{r^2+6\epsilon^2}{r^2+9\epsilon^2}.
\end{equation}

Before exhibiting the exact solution to this differential equation, let us study 
certain limit cases. For $r\gg\epsilon$, the equation reduces to \eqref{radialeq0}, 
so the general asymptotic solution is
\begin{equation}
\eqlabel{larger}
\text{R}^{m,l_1,l_2}(r,\epsilon)\sim
\frac{c_1}{r^2}J_{\sqrt{\Lambda^{m,l_1,l_2}+4}}(\sqrt{\lambda}r)+
\frac{c_2}{r^2}Y_{\sqrt{\Lambda^{m,l_1,l_2}+4}}(\sqrt{\lambda}r)\qquad(r\gg\epsilon),
\end{equation}
just like in the singular case. For $r\ll\epsilon$, \eqref{radialeq1} is also a 
Bessel equation, but the eigenvalues get shifted and the effective dimension of 
the base changes from $5$ to $3$. The normalizable 
solution at $r=0$ is
\begin{equation}
\eqlabel{smallr}
\text{R}^{m,l_1,l_2}(r,\epsilon)\sim\frac{1}{r}J_{\sqrt{\frac{2}{3}\Lambda_1^{m,l_1}
+4m^2+1}}\bigg(\sqrt{\frac{2}{3}\lambda-\frac{\Lambda_2}{9\epsilon^2}}r\bigg)
\qquad(r\ll\epsilon).
\end{equation}

The complexity of our problem lies in understanding how to choose $c_1,c_2$ (both 
real functions of $\epsilon,\lambda,l_1,l_2,m$) to patch up the two limiting 
behaviours. The solution for large $r$ eventually determines the spectrum upon 
evaluation at $r=r_\infty\gg\epsilon$ and it is not problematic in the limit 
$\epsilon\to 0$. However, the solution for small $r$ has $\epsilon$ in the 
denominator. One might thus suspect that the appropriate linear combination 
(\ie, the constants $c_1,c_2$) behaves uncontrollably in the limit. On the 
other hand, the validity range of the latter expansion is of shrinking size 
as $\epsilon\to 0$, so we cannot really progress much further with this 
approach.\footnote{Similar comments can be made by applying the Fuchs-Frobenius 
method on \eqref{radialeq1}. The recurrence relation determining the solution 
normalizable at $r=0$ in the vicinity of $r=0$ diverges in the limit 
$\epsilon\to 0$. Meanwhile, the corresponding series solution converges only 
within the radius excluding the nearest other singularity; that is in a disc 
whose size is controlled by $\epsilon^2$.}

To get a broader picture, we now examine the exact solution of \eqref{radialeq1} 
by recasting the equation in terms of some new independent and 
dependent variables:
\begin{equation}
\eqlabel{newvarx}
r\to x=-\frac{r^2}{9\epsilon^2}
\end{equation}
\begin{equation}
\eqlabel{newvary}
\text{R}^{m,l_1,l_2}(r,\epsilon)=x^{\frac{1}{2}\sqrt{\frac{2}{3}\Lambda_1^{m,l_1}+
4m^2+1}-\frac{1}{2}}(1-x)^{m/2}H^{m,l_1,l_2}(x,\epsilon).
\end{equation}
The multiplicative function we have inserted in \eqref{newvary} is non-vanishing 
on $[-\frac{r_\infty^2}{9\epsilon^2},0)$ and regular on the closure of this 
interval, which is the domain of $H^{m,l_1,l_2}(x,\epsilon)$. This function then 
still contains the full information about 
the spectrum. Substituting \eqref{newvarx}--\eqref{newvary} in \eqref{radialeq1} 
yields
\begin{equation}
\eqlabel{HeunC}
\frac{d^2H^{m,l_1,l_2}(x,\epsilon)}{dx^2}+\left(\frac{\gamma^{m,l_1}}{x}+
\frac{\delta^m}{x-1}-\beta\right)\frac{dH^{m,l_1,l_2}(x,\epsilon)}{dx}-
\frac{\alpha(\epsilon) x-q^{m,l_1,l_2}(\epsilon)}{x(x-1)}
H^{m,l_1,l_2}(x,\epsilon)=0
\end{equation}
where we have set
\begin{align}
\alpha(\epsilon)&=\frac{9\epsilon^2\lambda}{4}, \\
\beta&=0 \\
\eqlabel{gamma} \gamma^{m,l_1}&=1+\sqrt{\frac{2}{3}\Lambda_1^{m,l_1}+4m^2+1} \\
\delta^m&=m+1 \\
\eqlabel{q} q^{m,l_1,l_2}(\epsilon)&=\frac{3\epsilon^2\lambda}{2}+\frac{(m+1)}{2}
\bigg(1+\sqrt{\frac{2}{3}\Lambda_1^{m,l_1}+4m^2+1}\bigg)-
\frac{\Lambda_1^{m,l_1}}{12}-\frac{\Lambda_2^{m,l_2}}{4}-m^2-1
\end{align}

Eq.~\eqref{HeunC} is a standard form of the confluent Heun equation \cite{olver}, 
a degenerate version of the generic second order equation with four regular singular 
points (at $x=0, 1, a, \infty$). It is obtained from the general Heun equation by 
a ``confluence process'' (described e.g. in \cite{ronveaux}), which merges the 
singularity at $x=a$ with that at $x=\infty$. This results in a rank-2 irregular 
singular point at infinity and leaves behind the finite regular singular points 
at $0$ and $1$.\footnote{Actually, there is a restriction in the parameters of 
the general Heun equation which translates, after confluence, in the constraint 
$\alpha=\tilde{\alpha}\beta$, using the notations of \eqref{HeunC}. With this 
definition, $\beta$ cannot be set to zero independently of $\alpha$. Since this 
is precisely what we need here, we employ the slightly more general definition 
of the confluent Heun equation of \cite{olver} rather than the more widespread 
2.7.3 in ref. \cite{ronveaux}.} We will call $\text{HeunC}(\alpha,\beta,\gamma,
\delta,q;x)$ the solution of \eqref{HeunC} finite at $x=0$. It is normalized 
with the condition $\text{HeunC}(\alpha,\beta,\gamma,\delta,q;0)=1$. One can 
easily check that the $r\to 0$ behaviour of $\text{R}^{m,l_1,l_2}(r,\epsilon)$ 
(\cf~\eqref{newvary}) matches that derived from \eqref{smallr}.

The connection between Heun equation and the resolved conifold comes with a light 
sense of d\'ej\`a vu, as it occurred in related eigenvalue problems. For instance, 
Oota and Yasui \cite{oota} have exhibited its role in the spectrum of some  
five-dimensional toric Sasaki-Einstein manifolds (which include $T^{1,1}$ as a 
particular case). Eq. \eqref{HeunC} also comes about in the Laplacian eigenvalue 
problem on the Eguchi-Hanson space \cite{malmendier}. This space can be thought 
of as a resolution of a singularity on Calabi-Yau \textit{two}folds.

Further confluent cases of Heun equation can be obtained from \eqref{HeunC}, 
yielding so-called double-confluent, biconfluent, and triconfluent Heun equations. 
One might anticipate that solutions to confluent versions of an equation can be 
obtained as limit cases of solutions to the original equation. Although this is 
possible, one should worry about possible qualitative discrepancies due to the 
various different ways in which the two points can approach each other 
\cite{ronveaux}. For some cases, the spectra defined by the equations may 
be continuous \cite{veshev}; for others, drastic non-analytic changes can 
be expected \cite{lay,slavyanov}.

\begin{figure}[h]
\begin{center}
\includegraphics[scale=1]{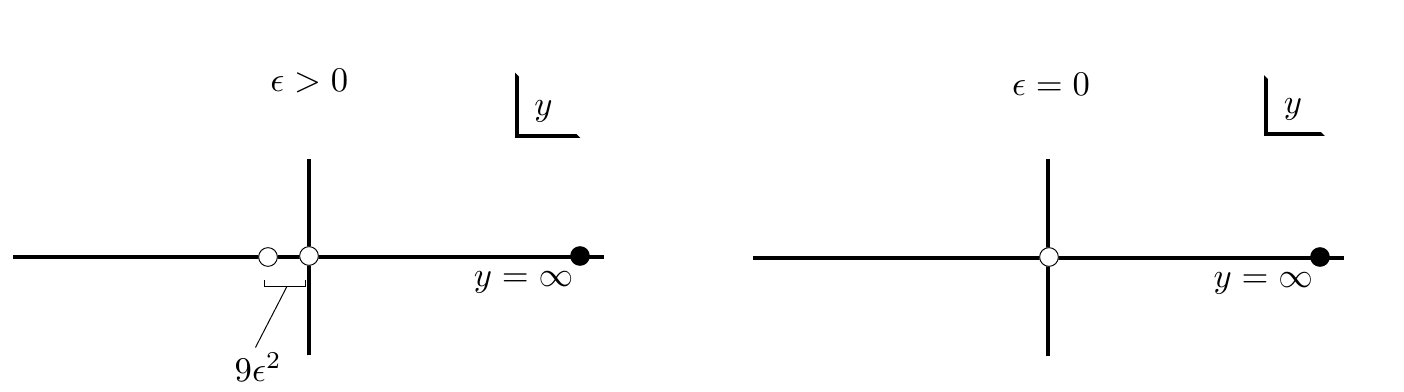}
\end{center}
\caption{Regular ($\circ$) and irregular ($\bullet$) singular points of 
\eqref{Heuntilde} in $\mathbb{C}_y$ for $\epsilon > 0$ (left) and $\epsilon=0$ (right).}
\label{fig3}
\end{figure}

These remarks take a critical importance regarding the question of convergence 
of the resolved conifold spectrum to that of the singular conifold. Indeed, the 
limit $\epsilon\to 0$ that we are interested in corresponds to yet another 
confluence process now taking the singular point at $1$ to that at $0$. This 
is better seen in terms of the variable $y=r^2=-9\epsilon^2x$:
\begin{equation}
\eqlabel{Heuntilde}
\frac{d^2H(y)}{dy^2}+\left(\frac{\gamma}{y}+\frac{\delta}{y+9\epsilon^2}+9
\epsilon^2\beta\right)\frac{dH(y)}{dy}+\frac{\lambda y/4 +q}{y(y+9\epsilon^2)}H(y)=0
\end{equation}
(dropping explicit labels). The singular points of this equation on the punctured 
Riemann sphere are pictured on fig.\ \ref{fig3} (left), while fig.\ \ref{fig3} (right) 
shows those for $\epsilon=0$.

The question whether the confluence process occurring here causes uncontrolled behaviour 
or not is much more delicate than it seems a priori, as discussed in the literature 
afore-cited. Also, the most readily available asymptotic expansions of confluent Heun 
solutions are ill-suited to study the connection between \eqref{larger} and 
\eqref{smallr} usefully. However, numerical tests strongly suggest that the transition
\begin{equation}
\frac{1}{r^2}J_{\sqrt{4+\Lambda^{m,l_1}}}(\sqrt{\lambda}r)\to\frac{1}{r^{2-\gamma^{m,l_1}}} 
(r^2+9\epsilon^2)^{m/2}\text{HeunC}\Big(\alpha(\epsilon),0,
\gamma^{m,l_1},\delta^m,q^{m,l_1,l_2}(\epsilon);-\frac{r^2}{9\epsilon^2}\Big)
\end{equation}
undergone by the eigenfunction upon resolution of the singularity is continuous and 
sufficiently well behaved for the heat trace to be bounded in the limit $\epsilon\to 
0$. We give, in the following section, an argument based on the full WKB expansion of 
\eqref{radialeq1}.\footnote{After this paper was largely completed, we came across ref. 
\cite{kazakov}.
The analysis of this paper, based on perturbation theory, is similar to our case but 
it does not directly apply because of the $\epsilon^2$-dependence of the parameter $q$ 
(\cf~\eqref{q}).}

\section{Full WKB Expansions of Radial Counting Functions}

Our argument is essentially the Bohr-Sommerfeld approximative quantization rule treated 
with some extra care. We thus need to recast the radial eigenvalue problem as a 
Schr\"odinger equation. Again, as a warm-up, we consider the singular case first.

In this section, we take advantage of our knowledge that the eigenvalues we are 
interested in arise from a collection of 1-d boundary value problems. In particular, 
instead of considering the full counting function 
\begin{equation}
N_{\hat{C}}(\lambda,\epsilon)=\sum_{l_1,l_2,m}n_{l_1,l_2,m}(\lambda,\epsilon),
\end{equation}
we focus on the \textit{individual} functions $n_{l_1,l_2,m}(\lambda,\epsilon)$, which 
count the number of eigenvalues---with  $l_1,l_2,m$ fixed---below $\lambda$. The sum 
is finite for any fixed value of $\lambda$ and the number of elements it contains is 
essentially bounded above by the volume in the space of quantum numbers, \ie, by Weyl's
law. We can thus concentrate on bounding each $n_{l_1,l_2,m}(\lambda,\epsilon)$ 
independently.

\subsection{Singular conifold}

Upon setting
\begin{equation}
\psi(r)=\psi^{m,l_1,l_2}(r)=r^{5/2}\text{R}^{m,l_1,l_2}(r),
\end{equation}
eq.~\eqref{radialeq0} becomes a Schr\"odinger equation:
\begin{equation}
\eqlabel{schro}
\hbar^2\frac{d^2}{dz^2}\psi(z)=Q(z)\psi(z),\qquad Q(z)=V(z)-\lambda,
\end{equation}
with $z=r$, $\hbar=1$, and the following radial confining potential:
\begin{equation}
\eqlabel{potentialsing}
V(z)=V^{m,l_1,l_2}(z,0)=\begin{cases}\frac{1}{z^2}\left(\Lambda^{m,l_1,l_2}+
\frac{15}{4}\right) & 0\leq z\leq r_\infty\\\infty & \text{else.}\end{cases}
\end{equation}

The WKB approximation to the bound states is obtained from the ansatz
\begin{equation}
\eqlabel{ansatz}
\psi(z)=\exp\left[\frac{1}{\hbar}\sum_{k=0}^\infty\hbar^kS_k(z)\right]
\end{equation}
by keeping only the first two terms in the series. Its validity increases as 
$\hbar\to 0$ or, for fixed $\hbar$ as $\lambda\to\infty$ \cite{bender}. The 
approximation is typically very precise on the whole range of the independent 
variable, except near the two turning points, defined by the condition $Q(z)=0$. 
Plugging \eqref{ansatz} into the Schr\"odinger equation yields
\begin{align}
\eqlabel{S0S1}
S_0'(z)&=\pm\sqrt{Q(z)}\\
S_1'(z)&=-\frac{1}{4}\ln Q(z).
\end{align}

Careful matching of the oscillatory behaviour in the region where $\lambda>V(z)$ 
(\ie,~$Q(z)<0$) with the exponential decay in the classically forbidden regions 
(beyond the turning points) yields the celebrated Bohr-Sommerfeld energy 
quantization rule\footnote{The shift of $1/2$ actually assumes that the potential 
has a finite slope at the turning points. For our model, the exact shift should be 
3/4, but we won't be picky about such details since we care only about the large $n$ 
limit.}:
\begin{equation}
\eqlabel{Bohr}
\left(n+\frac{1}{2}\right)\hbar\pi\approx\int\sqrt{-Q(z)}dz,\qquad(n=0,1,2,\ldots).
\end{equation}
The integral is taken from the smallest to the largest turning points.

Changing viewpoint, we can regard \eqref{Bohr} as an approximation to the individual 
counting functions introduced above. In the case of the singular conifold, this 
expression is found to describe very accurately the distribution of the zeros of 
the Bessel function. Carrying out the integral, with the potential \eqref{potentialsing}, 
gives
\begin{equation}
\eqlabel{ncones}
n_{l_1,l_2,m}(\lambda,0)\approx\frac{\sqrt{\mu^2-1/4}}{\pi}
\left[\sqrt{\frac{\lambda r_\infty^2}{\mu^2-1/4}-1}-\arccos
\left(\sqrt{\frac{\mu^2-1/4}{\lambda r_\infty^2}}\right)\right]-\frac{3}{4},
\end{equation}
where $\mu=\sqrt{\Lambda^{m,l_1,l_2}+4}$ is the order of the Bessel function giving 
rise to the spectrum. As a consistency check, it is possible to obtain precisely 
Weyl's law for any metric cone, with all proportionality factors, from this formula 
(assuming only that the law holds on the base manifold) \cf~appendix A. Ref.\
\cite{colin} also used a minor variant of this formula to obtain 
the second term of the expansion in the case of the 2-dimensional disc.

\subsection{Resolved conifold}

In the case of the resolved conifold, eq. \eqref{radialeq1} becomes a Schr\"odinger 
equation at the expense of a complicated change of variable necessary to remove the 
factor multiplying the eigenvalues:
\begin{equation}
r\to z=-i\sqrt{6}\epsilon \text{E}\Big(\frac{ir}{3\epsilon},\frac{\sqrt{6}}{2}\Big).
\end{equation}
It arises from the property
\begin{equation}
\eqlabel{z}
\frac{dz}{dr}=\sqrt{\frac{r^2+6\epsilon^2}{r^2+9\epsilon^2}}.
\end{equation}
Here $\text{E}(x,k)$ denotes the incomplete elliptic integral of the second kind. 
Following this change of variable, we must substitute the dependent variable as
\begin{equation}
\psi(z)=\psi^{m,l_1,l_2}(z,\epsilon)=\sqrt{r^3\sqrt{(r^2+6\epsilon^2)(r^2+9\epsilon^2)}}
\text{R}^{m,l_1,l_2}(z,\epsilon)
\end{equation}
to obtain an equation in the form \eqref{schro}. The effective potential energy is 
conveniently written in terms of $r$, which should now be regarded as a function of 
$z$:
\begin{equation}
\eqlabel{potentialres}
V^{m,l_1,l_2}(z,\epsilon)=\frac{\Lambda^{m,l_1,l_2}(r,\epsilon)}{r^2}+
\frac{15r^8/4+90r^6\epsilon^2+765r^4\epsilon^4+2592r^2\epsilon^6+2187
\epsilon^8}{r^2(r^2+6\epsilon^2)^2(r^2+9\epsilon^2)^2}\left(\frac{dz}{dr}\right)^{-2}.
\end{equation}
(Of course, the potential is again infinite outside the range $0\leq r<r_\infty$.)

In the present case, the Bohr-Sommerfeld integral cannot be performed exactly as 
previously. We observe instead that the potential function reduces to 
\eqref{potentialsing} for $\epsilon=0$ (thus the choice of notations). Hence, 
the left turning point of the resolved problem limits to that of the singular 
problem, that is $\sqrt{(\Lambda^{m,l_1,l_2}+15/4)/\lambda}$, when $\epsilon\to 
0$. Since this is always strictly positive (albeit small), no divergence can be 
due to evaluation of \eqref{Bohr} at the turning points. This holds regardless 
of the value of $\epsilon$.

If any discontinuity occurs in the confluence process, we should thus see its 
effects in the integrand only. Nothing of this kind occurs in the leading Bohr 
term as can be seen easily from the potential function \eqref{potentialres}. 
However, given the asymptotic analysis done on the heat trace in section 5, we 
could expect negative powers of $\epsilon$ to arise at higher orders.

Let us then consider the full formal solution of Schr\"odinger equation in terms 
of the WKB ansatz \eqref{ansatz}. The next terms after \eqref{S0S1} can be obtained 
recursively from \cite{bender}
\begin{equation}
\eqlabel{recurrence}
2S'_0(z)S'_k+S''_{k-1}(z)+\sum_{j=1}^{k-1}S'_jS'_{k-j}=0,\qquad(k=2,3,4,\ldots).
\end{equation}
The leading of these corrections are
\begin{align}
S'_2(z)&=\pm\left[\frac{Q''(z)}{8Q(z)^{3/2}}-\frac{5(Q'(z))^2}{32Q(z)^{5/2}}\right],\\
S'_3(z)&=-\frac{Q''(z)}{16Q^2(z)}+\frac{5(Q'(z))^2}{64Q^3(z)},\\
S'_4(z)&=\pm\left[\frac{Q''''}{32Q^{5/2}}-\frac{7Q'Q'''}{32Q^{7/2}}-
\frac{19(Q'')^2}{128Q^{7/2}}+\frac{221Q''(Q')^2}{256Q^{9/2}}-
\frac{1105(Q')^4}{2048Q^{11/2}}\right].
\end{align}

It is easy to convince oneself that the structure of these expressions is qualitatively 
the same at any order: sums of ratios of derivatives of the shifted potential $Q(z)$ 
to powers of $Q(z)$. It is a known fact that odd terms are total derivatives and 
single-valued. The even terms all involve square roots of $Q(z)$. Thought of as 
complex functions, they have two branches (thus the $\pm$ signs). \cite{bender}

The two independent solutions obtained by summing all these terms are exact but 
generally divergent on the whole complex $z$ plane. They must be interpreted as 
asymptotic expansions of the true solutions. The Bohr-Sommerfeld rule is 
consequently understood as the first term of the full WKB expansion \cite{dunham,bender}
\begin{equation}
\eqlabel{fullWKB}
\left(n+\frac{1}{2}\right)\sim\frac{1}{2\pi i}\oint\frac{1}{ \hbar}
\sum_{j=0}^\infty\hbar^{2j}S'_{2j}(z)dz,\qquad \lambda\to\infty.
\end{equation}
The contour encircles the two turning points, and no other singularity in the 
complex $z$ plane. The sign of the $S'_{2j}(z)$ must be adjusted so that the 
integral is performed on a smooth branch whose cut connects the turning points 
on the real axis.

We now make our argument on boundedness of the counting functions 
$n_{l_1,l_2,m}(\lambda,\epsilon)$ and thus, by our previous discussion, of 
$Z_{\hat{C}}(t,\epsilon)$. We again identify $n$ in \eqref{fullWKB} as an 
individual counting function (corresponding to the quantum numbers $l_1,l_2,m$).

Unlike the Minakshisundaram-Pleijel heat trace expansion (\cf~\eqref{expansion}), 
the full WKB expansion does \textit{not} exhibit any divergence in the limit 
$\epsilon\to 0$. This can be seen after close inspection of \eqref{recurrence}, 
\eqref{potentialres} and \eqref{z}, and indicates that the singular space 
heat trace $Z_{C}(t)=Z_{\hat{C}}(t,0)$ can be used as a starting point to bound 
$Z_{\hat{C}}(t,\epsilon)$.

One may reasonably ask why we should trust the continuity of the WKB expansion more 
than the divergence of the heat trace expansion? As was discussed in section 5, 
the latter expansion is in fact ill-suited to our discussion as it is valid for 
$t\to 0$ and $\epsilon$ fixed, while we need the opposite regime. Also, and this 
is the key remark, its qualitative form changes discontinuously when 
$\epsilon=0$ (\cf~subsection 5.4). \textit{This is not the case for the WKB 
expansion.} Indeed, we saw in subsection 7.1 that WKB ideas \textit{can} be usefully 
exploited to solve non-trivial questions even in the singular case. The qualitative 
form of the expansion is insensitive to the existence (or not) of a singularity. 
We conclude therefore that the conclusions drawn from \eqref{fullWKB} are in 
fact correct, and maintain that $Z_{\hat{C}}(t_*,\epsilon)$ can be bounded above 
independently of $\epsilon$ in the limit $\epsilon\to 0$.

\section{Discussion}

We have argued in the previous section that the spectrum of the scalar Laplacian
on the resolved conifold is continuous in the singular limit, as expected on
general mathematical grounds. The main step in the argument is based on the 
analysis of the 1-dimensional radial Schr\"odinger equation that results after 
separation of variables. The divergence of the asymptotic expansion of the 
trace of the heat kernel is a remnant of the breakdown of the classical 
propagation across the singularity, while the quantum mechanical evolution 
is well-behaved at finite energy. This confirmation of physical intuition makes 
us confident that analogous statements should hold for more general singularities 
as well. An obvious next test case would be the deformed conifold, on which the
wave equation is not reducible to a purely one-dimensional problem because
of the smaller symmetry algebra.

We now wish to discuss the lessons for the question \eqref{question} that
we proposed as a geometric supplement to the SCHOK bound \eqref{schok} on the 
number of marginal deformations of a conformal field theory. 

Since we are allowed to move around in the moduli space of 
Ricci-flat metrics, we can imagine approaching the question by speculating 
that a general 
compact Calabi-Yau manifold of complicated topology can, by deformation of 
the metric, be decomposed into flatter regions that are connected and/or
terminated by regions of concentrated curvature, intuitively similar
to what is possible for Riemann surfaces. The question is then whether
the contribution to the heat trace from the curved regions can be
estimated and bounded in a uniform fashion.

Now, had it turned out that the heat trace were in fact {\it not} continuous
at the approach of a singularity, we would have had to stay at a finite
distance from all singularities in the moduli space of metrics. Since we
expect the number of singularities to increase with the dimension of the
moduli space, it would be very delicate to find a region in which to 
bound the heat trace in very high-dimensional moduli spaces, and any
resulting bound would unlikely be uniform.

Under the hypothesis of spectral continuity however, curvature singularities 
(at least those of the type we analyzed) do not in fact preclude a bound on 
the heat trace. On the contrary, if the manifold can in fact be simplified 
in the way described above, the singular limit could prove a useful starting 
point from which to estimate the heat trace on the smooth manifold.
This has a chance of being uniform in the topology and the string scale.

\begin{acknowledgments}
We thank Simeon Hellerman for asking the question which led to this research,
as well as for collaboration at an initial stage. We would like to thank 
Dmitry Jakobson, Fr\'ed\'eric Rochon, and David Sher for valuable discussions
related to the heat kernel expansion. This research was supported by an Alexander 
Graham Bell Canadian Graduate Scholarship (NSERC), a Bourse de ma\^itrise en 
recherche (FRQNT), funds provided by a Tier II Canada Research Chair, and an 
NSERC Discovery Grant.
\end{acknowledgments}

\appendix

\section{Weyl's law for cones from WKB expansion}

As an application of formula \eqref{ncones}, we obtain here Weyl's law for any 
$(d+1)$-dimensional metric cone $C_B$ over a $d$-dimensional manifold $B$ 
($ds_{C_B}^2=dr^2+r^2ds_B^2$). As usual, we assume the cone is truncated at 
$r=r_\infty$. We use only the differential equation: the exact solution in 
terms of Bessel functions is unnecessary. We also assume that Weyl's law is 
satisfied on the base, that is, if $\Lambda$ and $N_B(\Lambda)$ are respectively 
the eigenvalues and the counting function associated with the Laplacian $\Delta_B$ 
on the base, we have
\begin{equation}
\eqlabel{weylapp}
N_B(\Lambda)\approx\left(\frac{\Lambda}{4\pi}\right)^{d/2}\frac{\vol(B)}{\Gamma(d/2+1)},
\qquad (\Lambda\to\infty)
\end{equation}
The goal is to derive this formula with $B$ replaced by $C_B$, $\Lambda$ by $\lambda$ 
(the eigenvalue of the full problem), and $d$ by $d+1$. This form of Weyl's law is 
related to \eqref{weyl} through a Laplace transformation.

As discussed in the main text (\cf~\eqref{radialeq0}), the eigenvalue problem for 
the Laplacian $\Delta_{C_B}$ reduces to the single differential equation
\begin{equation}
\eqlabel{conerad}
\frac{1}{r^d}\frac{d}{dr}\left(r^d\frac{d}{dr}R(r)\right)+\left(\lambda-
\frac{\Lambda}{r^2}\right)R(r)=0.
\end{equation}
The WKB method applied on this equation (rewritten in Schr\"odinger form) yields the 
quantization condition \eqref{ncones}:
\begin{equation}
\eqlabel{approxzeros}
n\approx\frac{1}{\pi}
\left[\sqrt{\lambda r_\infty^2-\mu^2+1/4}-\sqrt{\mu^2-1/4}\arccos
\left(\sqrt{\frac{\mu^2-1/4}{\lambda r_\infty^2}}\right)\right]-\frac{3}{4}.
\end{equation}

Here, $\mu=\sqrt{\Lambda+(d-1)^2/4}$ is the order of the Bessel equation in which 
\eqref{conerad} can be cast. We can regard this as an implicit expression of $\lambda$ 
as a function of the radial quantum number $n$ and the base space quantum numbers 
(encapsulated in $\Lambda$). Fixing a continuous value for $\lambda$, \eqref{approxzeros} 
can alternatively be regarded as a function of $\Lambda$ delimiting the region of phase 
space with energies below $\lambda$. To leading order, $N_{C_B}(\lambda)$ is then given 
by the volume of this region:
\begin{equation}
\eqlabel{NCB}
N_{C_B}(\lambda)\approx\int n(\Lambda)\big|_{\lambda}dN_{B}=\int n(\Lambda)
\big|_{\lambda}\frac{dN_{B}(\Lambda)}{d\Lambda}d\Lambda.
\end{equation}
The range of integration is determined as follows. Intuitively, the smallest positive 
value $\Lambda$ can take is reached when $n(\Lambda)\big|_{\lambda}$ is maximal (for 
fixed $\lambda$). This happens roughly when $\mu^2-1/4=0$. Conversely, the upper limit 
is obtained from the minimal (positive) value of $n(\Lambda)\big|_{\lambda}$. It will 
turn out to be sufficiently precise to consider this to happen when 
$\mu^2-1/4=\lambda r_\infty^2$.

Defining the integration variable $x=\sqrt{\frac{\mu^2-1/4}{\lambda r_\infty^2}}$, 
\eqref{weylapp} gives (to leading order)
\begin{equation}
\frac{dN_B(\Lambda)}{d\Lambda}\approx(x\sqrt{\lambda}r_\infty)^{d-2}
\frac{d/2}{(4\pi)^{d/2}}\frac{\vol(B)}{\Gamma(d/2+1)}
\end{equation}
and \eqref{NCB} becomes (dropping the shift by $3/4$)
\begin{align}
N_{C_B}(\lambda)&\approx\int_0^1\frac{\sqrt{\lambda}r_\infty}{\pi}\left[\sqrt{1-x^2}-x
\arccos{x}\right](\sqrt{\lambda}r_\infty)^{d}\frac{d}{(4\pi)^{d/2}}
\frac{\vol(B)}{\Gamma(d/2+1)}x^{d-1}dx \nonumber \\
&=\frac{(\sqrt{\lambda}r_\infty)^{d+1}}{\pi}\frac{d}{(4\pi)^{d/2}}
\frac{\vol(B)}{\Gamma(d/2+1)}\int_0^1\left[\sqrt{1-x^2}-x\arccos{x}\right]x^{d-1}dx.
\end{align}

We immediately see that the power of $\lambda$ is as expected. The integral can be 
worked out analytically. It is straightforward to verify from here that the factors 
match Weyl's law exactly. The volume arises as $\vol(C_{B})=r_\infty^{d+1}\vol(B)/(d+1)$.

\end{document}